\documentclass[
    ,final            
  ]
  {aipproc}

\layoutstyle{8x11double}

\def\la{\hbox{\rlap{\raise.3ex\hbox{$<$}}\lower.8ex\hbox{$\sim$}\ }}
\def\ga{\hbox{\rlap{\raise.3ex\hbox{$>$}}\lower.8ex\hbox{$\sim$}\ }}

\def\ni{\noindent}

\def\about{$\sim$}
\def\arcsec{$\,^{\prime\prime}$}

\def\deg{$^{\circ}$}
\def\erg/cm2sec{ergs~cm$^{-2}$~s$^{-1}$}  
\def\ergcm2{ergs~cm$^{-2}$}  
  
\def\mdot{$\dot{m}$~}  
\def\X{$\times$~}
\def\Fx{F$_x$~}

\def\Lx{L$_x$~}

\def\pc3{pc$^{-3}$~}

\def\apj{ApJ}
\def\aap{A\&A}

\newcommand{\lsim }{{\lower0.8ex\hbox{$\buildrel <\over\sim$}}}
\newcommand{\gsim }{{\lower0.8ex\hbox{$\buildrel >\over\sim$}}}

\newcommand{\Msun}{\ifmmode {M_{\odot}}\else${M_{\odot}}$\fi~}
\newcommand{\Rsun}{\ifmmode {R_{\odot}}\else${R_{\odot}}$\fi}
\newcommand{\Lsun}{\ifmmode {L_{\odot}}\else${L_{\odot}}$\fi}
\newcommand{\mv}{\ifmmode {m_{V}}\else${m_{V}}$\fi}
\newcommand{\Mv}{\ifmmode {M_{V}}\else${M_{V}}$\fi}
\newcommand{\lopt}{\ifmmode L_{opt} \else $~L_{opt}$\fi}
\newcommand{\loglopt}{\ifmmode{\rm log}~L_{opt} \else log$~L_{opt}$\fi}
\newcommand{\lx}{\ifmmode L_x \else $~L_x$\fi}
\newcommand{\loglx}{\ifmmode{\rm log}~L_x \else log$~L_x$\fi}
\newcommand{\cmsq}{\ifmmode{\rm ~cm^{-2}} \else cm$^{-2}$\fi}
\newcommand{\nh}{\ifmmode{\rm N_{H}} \else N$_{H}$\fi}
\newcommand{\fcgs}{\ifmmode {\rm erg~cm}^{-2}~{\rm s}^{-1}\else
erg~cm$^{-2}$~s$^{-1}$\fi} 
\newcommand{\lcgs}{\ifmmode erg~~s^{-1}\else erg~s$^{-1}$\fi}

\begin{document}

\title{Hard X-ray Timing with EXIST}

\author{Jonathan E. Grindlay}{
  address={Harvard-Smithsonian Center for Astrophysics, 60 Garden St.,
  Cambridge, MA 02138}
}

\begin{abstract}
The Energetic X-ray Timing Survey Telescope (EXIST) mission concept is  
under study as the Black Hole Finder Probe (BHFP), one of the three 
Einstein Probe missions in the Beyond Einstein Program in the current 
NASA Strategic Plan. EXIST would conduct an all-sky imaging hard 
X-ray ($\sim$10-600 keV) survey with unprecedented sensitivity: about 
5 $\times 10^{-13}$ cgs over any factor of 2 bandwidth, or comparable to that 
achieved at soft X-rays in the ROSAT survey. The proposed angular 
resolution of 5arcmin, temporal resolution of 10microsec, energy 
resolution of 1-4 keV over the broad band, and duty cycle of 0.2-0.5 for 
continuous coverage of any source provide an unprecedented  
phase space for timing and spectral studies of black holes --from 
stellar to supermassive, as well as neutron stars and accreting white dwarfs. 
The large sky coverage allows intrinsically rare events to be studied. 
One particularly exciting example is the possible detection of tidal 
disruption of stars near quiescent AGN. Super flares from 
SGRs could be detected out to the Virgo cluster. The large duty cycle 
and all sky monitor nature of the mission will enable 
QPOs from luminous AGN and BH X-ray binaries to be studied 
on timescales not possible before. I provide an overview of the 
mission concept and Reference Design, the X-ray timing science 
prospects for EXIST, and how these might be further optimized  
in the current Study for EXIST as the BHFP so that EXIST might  
include many of the desirable features of a next-generation timing 
mission.  
\end{abstract}
\maketitle


\section{Introduction}
At energies above $\sim$10 keV, the sky becomes relatively dark as the 
thermal emission from the brightest X-ray binaries becomes less 
prominent and diffuse emission from hot gas can not be 
maintained. The number of bright sources decreases dramatically, but 
those that remain are likely to be among the most interesting (at 
least for high energy astrophysics) as they are either accretion 
powered or non-thermal. In both cases, variability often becomes 
increasingly prominent with increasing energy, at least up to 20-30 
keV beyond which it has usually been difficult to measure (on 
short timescales) for persistent sources. These hard X-ray sources 
are, in general, accreting black holes: at fluxes \Fx(10-20 keV)  
\la1 \X 10$^{-11}$ \fcgs, the extragalactic AGN dominate the sky 
above 10 keV. But even near the galactic plane, the logN-logS relations 
derived from the RXTE/ASM data for \Fx(2-12keV) \ga 6 \X \X 10$^{-11}$ \fcgs 
by Grimm, Gilfanov and Sunyaev (2002) (and discussed by 
Vrtilek in this volume) would suggest that black holes are \ga10\% of 
the total for source fluxes at these limits, and 
probably more so if instead measured at 10-20 keV. 
This inference requires including transients, and the simple fact 
that of the \about180 galactic sources detected above the limiting \Fx 
by the ASM, all 16 of the currently known black hole candidates 
(BHCs) in the Galaxy (cf. McClintock and Remillard 2003) would have 
been detected. 

It could then be argued that a follow-on mission to RXTE, with its 
highly successful timing studies of accreting black 
hole (BH) and neutron star (NS) systems, would gain most from 
greatly increased sensitivity 
above 10 keV. Obviously, this is countered by the greater source 
fluxes and thus count rates at lower energies: for a simple power 
law (with photon index 2) approximation to the spectrum, the counts 
are 10\X larger in a decade lower energy band. However the relatively 
non-variable source emission components (e.g. blackbody emission 
from NS surfaces or the thermal soft 
continuum from accretion disks around BHs) 
provide additional `signal backgrounds' at low energies, 
and indeed the rms amplitude of quasi-periodic 
oscillations (QPOs) in both BH and NS accretors increases sharply 
with energy up to at least \about20 keV (cf. Kaaret, this volume). 
Thus on balance, the net gains with increased energy band are 
dependent on the timescale and spectrum of the variable source 
component(s) to be measured. QPOs can be detected in time T with  
detection significance 
n$_{\sigma}$  (van der Klis 1998) 

\begin{equation}
T = 4\frac{n_\sigma^2\Delta\nu}{r^4s^4}(b+s)^2
\end{equation}

\ni
where s is the source count rate, b is the background, $\Delta\nu$ 
is the signal bandwidth, and r is the rms amplitude. 

We shall use this expression to estimate the approximate 
sensitivity  for QPOs that could be detected at hard X-ray 
energies with the Energetic X-ray Imaging Survey Telescope (EXIST), 
now under Study as the proposed Black Hole Finder 
Probe (one of the 3 Einstein Probe missions) in the Beyond 
Einstein Program under the current NASA Strategic Plan. 
In making these estimates, and others for hard X-ray timing science 
possible with EXIST, we make use of the so-called ``Reference Design'' 
for the mission. We first introduce the mission concept and 
Reference Design and then discuss the principal areas of (hard) 
X-ray Timing that could be advanced with EXIST. These include:  
QPOs for bright galactic sources and their long-term trends, QPOs for 
AGN, stellar encounter events (both stellar disruption and 
star-disk interactions) with dormant AGN, and super-flare 
events from soft gamma-ray repeaters (SGRs) in galaxies out 
to the Virgo cluster. We then describe some of the modifications 
to the Reference Design which are being investigated in the current 
BHFP Study and which could further optimize EXIST as a next generation 
timing mission.

\section{EXIST mission concept: Reference Design}

\subsection{Capsule history of EXIST concept}
The EXIST mission has been under study, in several phases, over the 
past decade. Initially proposed as a MIDEX (Intermediate Explorer) 
mission (Grindlay et al 1995), it was later studied for a much 
more sensitive (and larger) configuration that could achieve all-sky 
imaging survey sensitivities at 10-100 keV comparable to what the only  
all-sky imaging soft X-ray survey, done with ROSAT in 
1990-91 (Voges et al 1999), had 
achieved at 0.1-2.5 keV. The greatly enhanced survey sensitivity 
merited the endorsement of EXIST in the Decadal Survey 
Report\footnote{{\it Astronomy and Astrophysics in the New 
Millennium}, National Academy Press (2001)} as 
one of the three high energy astrophysics missions recommended for 
the coming decade (the other two being Constellation-X and GLAST). 
The ``larger'' EXIST was first studied (Grindlay et al 2001) 
for possible implementation on the International Space Station 
(given its large mass and nominal zenith pointing requirement). 
However, complexities of planning or (even) accomodating for 
large attached payloads on ISS made it clear that a lower real cost 
implementation could be achieved as a free flyer, which could also 
be more sensitive (given the significantly lower particle 
backgrounds available in lower inclination, and altitude orbits). 
Thus EXIST was studied as a free flyer (low earth orbit; range of 
inclinations) at the Integrated Mission Design Center (IMDC) 
at the NASA Goddard Space Flight Center in 2001 with additional 
studies throughout 2002. The resulting design for a free flyer 
EXIST, the Reference Design (Grindlay et al 2003a), was the basis 
for our proposal that a sensitive hard X-ray survey be considered 
for the ``Beyond Einstein Program'' which was being formulated 
in 2002 by the advisory committee (SEUS) for the Structure and 
Evolution of the Universe (SEU) Theme of NASA's Office of Space 
Science. The Beyond Einstein Program, as finally announced 
in early 2003 and embodied in the current NASA Strategic Plan, 
includes the Black Hole Finder Probe (BHFP), one of the three 
Einstein Probe missions proposed to be conducted in addition 
to the two primary missions (LISA and Constellation-X) 
during the  Beyond Einstein Program. 
EXIST is now beginning a Study (as one of two concepts for BHFP) 
for how the BHFP mission should be defined so that a subsequent 
competition for its actual development and implementation can 
be carried out.  

\subsection{Summary of the Reference Design}
As a hard X-ray survey mission with primary emphasis on black 
hole surveys in both space and time, EXIST maximizes the two 
primary determinants for sensitivity: total detection area 
and total exposure time, each as appropriate for any given source. 
This means that the mission design must incorporate very 
wide field of view (FoV) telescopes and detectors, which in turn 
means that grazing incidence optics (e.g. with multilayers) are 
precluded. The basic concept, then, is a scanning array of coded 
aperture imaging telescopes, each with 60\deg~ \X 75\deg~ fully-coded 
FoV. The three together combine to form a fully-coded fan beam 
180\deg~ \X 75\deg, with partial coding for imaging (with some loss 
of sensitivity) out to even larger total solid angle of approximately 
5sr (FWHM). The 180\deg~ axis of the combined beam is oriented 
perpendicular to the orbital ram direction so that the full sky is 
swept out each 95min orbit, and each source on the orbital equator 
is observed for at least 75/360 $\sim$21\% of the total time at full 
sensitivity (except for those $\sim$5 orbits out of 15 each day where 
the satellite's passage near the South Atlantic Anomaly (SAA) would 
cut perhaps 20\% of the exposure on those orbits for a nominal 28\deg~ 
inclination orbit. This survey time loss, amounting to perhaps 
15\%, could be reduced considerably by launching into a lower 
inclination orbit -- at the cost of a larger launch vehicle or 
smaller payload mass.  

The Reference Design telescope and spacecraft structure has the 
three large area telescopes, each with 2.7m$^2$ 
total area of imaging Cd-Zn-Te (CZT) detectors  
(each fabricated with a 16 \X 16 array of 1.2mm pixels 
evaporated on the anode side of 2cm \X 2cm \X 0.5cm crystals, and 
read out by a single ASIC direct-bonded to each crystal),  
and still larger area (factor of $\sim$2) passive coded aperture 
masks at focal length 1.5m above each telescope, ``stacked'' 
on the spacecraft (S/C) structure below. This is the minimum mass 
and envelope configuration for the desired total detector area
(8m$^2$) required to achieve the desired survey flux sensitivity  
of \about5 \X 10$^{-13}$ \fcgs (comparable to that at soft 
X-ray energies achieved with ROSAT) in any factor of 2 band of energy, 
from \about10-200 keV, over a 6-12month total exposure time for 
the fields of view (and thus exposure time per orbit) given. 
Due to increasing transparencies of the CZT detectors, shields, 
and coded masks, the sensitivity becomes \about10\X lower over  
the band \about200-600 keV. A view 
of the EXIST Reference Design (hereafter EXIST-RD), with 
solar panels deployed and oriented relative to its orbital vector, is shown as 
Figure 1 in Grindlay et al (2003a). Here we show (Fig. 1) a more 
schematic view of just the 3-telescope stack on the S/C, with a 
scale figure along side for comparison.

\begin{figure}
\includegraphics[height=.28\textheight]{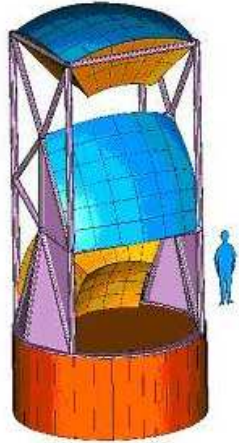}
\caption{Schematic view of EXIST: 3 telescopes, each with curved 
coded aperture mask 1.5m above its curved detector plane, 
all mounted above the spacecraft bus.}
\end{figure}

We also show (Fig. 2) two side views of the telescope-S/C stack (the 
right view makes the 180\deg~ fan beam FoV of the three 
telescopes combined easier to envision), 
and finally (Fig. 3) the configuration in the Delta IV fairing for 
launch. A summary of some of the key  parameters for the 
EXIST-RD is given in Table 1.

\begin{figure}
\includegraphics[height=0.28\textheight]{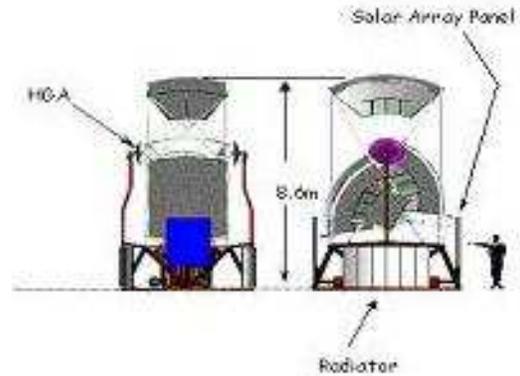}
\caption{EXIST with antennas and solar panels folded for launch.}
\end{figure}

\begin{figure}
\includegraphics[height=0.28\textheight]{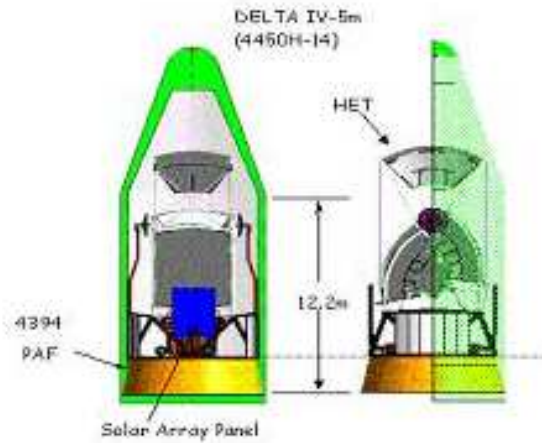}
\caption{EXIST in Delta IV shroud.}
\end{figure}


\begin{table}
\begin{tabular}{lr}
\hline
\tablehead{1}{l}{b}{Parameter}
  & \tablehead{1}{r}{b}{Value or range} \\
\hline
Energy range & 10-600 keV \\
 & (\la10 -- \ga300 keV) \\
Sens. (5$\sigma$) & 50$\mu$Crab\tablenote{50$\mu$Crab= 5 \X10$^{-13}$
 \fcgs in band $\Delta$E=E-0.5E}~ (10-200 keV)\\
& 500$\mu$Crab (200-600 keV) \\
FoV & 180\deg \X 75\deg~ (fully coded) \\
Survey coverage & full sky ea. orbit \\
Angular resol. & 3-5' \\
Source loc. & 10-50'' \\
Energy resol. & 1-4 keV \\
Temporal resol. & 10$\mu$sec \\
Detectors & 8m$^2$ imaging Cd-Zn-Te \\
& (1.2mm pixels on 2cm crytals) \\
Telescopes & 3, each coded aperture \\ 
& (URAs, 2.5mm pixels) \\
Mission ops. & zen. scan; \& pointings \\ 
Mass, power, TM & \about 8000kg, 1500W, 2Mbs \\ 
\hline
\end{tabular}
\caption{EXIST Reference Design key parameters}
\label{tab:a}
\end{table}

\section{Key Timing Science with EXIST}
\subsection{EXIST as a component of an XRT mission?}
With its heretofore unique capability of all-sky imaging each 
95min (or nearly so), and high sensitivity above 10 keV, EXIST would 
enable a number of new (or complementary) temporal-spectral surveys 
not yet carried out. The characteristics of the EXIST-RD  
include at least two, if not three, of the widely-agreed 
upon requirements for a follow-on X-ray Timing (XRT) mission to 
succeed RXTE: \\

\ni{\bf Large area:} For the EXIST-RD, any given source 
is imaged by \about1.5m$^2$ at any given time, although only 
half this area is exposed to the source (the other half, occulted 
by the half-open coded mask, is measuring background simultaneously as well 
as all other sources in that instantaneous FoV). 
Thus for fast timing studies (e.g. kHz QPOs, etc.), 
EXIST-RD is only comparable in area, and thus source count 
rate, to RXTE at 10 keV. At \about20-30 keV, however, where QPOs 
are typically a factor of 2-3 larger in rms amplitude (e.g., Kaaret
2004), and the RXTE detector efficiency has fallen by a factor of 
(roughly 4) while EXIST-RD has maintained (nearly) unit QE, the 
increase in signal or effective area is closer to the desired factor of
\about10. The effects of the higher detector background for 
EXIST-RD, with its very large FoV, are discussed below. \\

\ni{\bf ASM capability:} There is wide agreement that a 
proper timing mission must include an all sky monitor (ASM) 
capability as part of its mission plan. The most interesting 
phenomena, which may be the most rare, will otherwise surely be 
(usually) missed. Here the EXIST-RD is in its element, with an ASM 
sensitivity (\about500$\mu$Crab per day) that is a factor of 
10-100\X more sensitive than the RXTE/ASM (2-12keV) and 
\ga100\X more sensitive than the ASM ability achieved by 
occultation imaging with BATSE, which resulted in the most 
sensitive hard X-ray (20-100 keV) detection, \about1\deg  
locations, and temporal variability studies achieved to date 
(cf. Harmon et al 2004, Shaw et al 2004). \\

\ni{\bf Hard X-ray response:} Given the energy dependence 
of QPOs just mentioned, as well as the general arguments in the 
Introduction above, it is clear that temporal variability is often 
(but not always; see below) best studied above \about10 keV. 
Here again the EXIST-RD gains on low energy optimized (e.g. 
focusing optics) approaches to very large areas. True, a 
grazing incidence large area optic has negligible background in 
comparison to a large area coded aperture HX imager, but as 
discussed below this is not an overwhelming advantage if the source(s) 
being observed (for kHz QPOs, say) are bright: if total background 
rates, b, are comparable to source count rate s (cf. Eq. 1), then 
the required integration time is ``only'' increased by a factor of 
\about1.4. 


For bright galactic sources (e.g. the microquasar GRS1915+105), 
EXIST-RD can measure the energy dependence of fluctuation 
amplitudes over a wider range of energies extending well beyond 
the (typical) 15-20 keV cutoff of the PCA. The nominal background 
expected in the wide FoV (50\deg \X 60\deg~ in each detector 
module) seen by each CZT detector element is dominated by 
the diffuse cosmic flux and is comparable to 
the Crab flux in the \about20-40 keV band and decreasing more 
steeply at higher energies. For an increase 
in effective area of the EXIST-RD vs. the PCA at 20-40 keV estimated 
as a factor of \about10, and correspondingly the increase in signal 
count rate, s, but now with the non-source background b \about s, 
Eq. (1) suggests EXIST-RD would detect QPOs at these energies in only 
\about4\% of the time as with the PCA. It is already clear that 
GRS1915+105 has strong QPOs with  rms amplitude increasing with 
energy up to (at least) 20-30 keV (Rodriquez et al 2002).  
EXIST-RD would push these studies to higher energies and allow 
their continuous monitoring (every orbit) for new constraints 
on the underlying physics of QPOs from stellar mass BHs.  

It is also clear that EXIST-RD is competitive, as a 
timing mission, for temporal  
measurements of relatively faint sources on long timescales. 
This regime for QPOs is of course most relevant for AGN and is 
thus far only studied with the limited coverage of long 
campaigns of PCA pointings on a half dozen moderately bright AGN 
(Markowitz et al 2003).  
EXIST-RD could measure, continuously,  
AGN fluctuations on minimum 
timescales of $\delta\tau_{min}$ \about0.5-1day (depending 
on source orbital latitude) at the 
\about500$\mu$Crab level. For a 10$^8$ \Msun BH, this 
enables detection of QPOs down to the plausible upper 
frequency cutoffs expected for disks extending down to 
3R$_{Schwarz}$, for which $\tau_{min}$ \about 2.4d is expected. 
In fact the PCA monitoring campaigns of Markowitz et al (2003) 
have obtained interesting evidence for high frequency breaks 
in the power spectra of several of the 6 AGN monitored which 
are consistent with a scaling $\tau_{min,days}$ 
\about M$_{BH}$/10$^{6.5}$ \Msun which -- taken literally -- 
would suggest the (predominantly) 2-10 keV X-ray emission 
observed with the PCA is arising from regions at 
\about15R$_{Schwarz}$. Comparison with similar power spectra 
obtained on these objects (and many more) with EXIST would 
constrain the origin of variability and the evolution of 
$\tau_{min}$ with source luminosity and accretion rate 
variations. The  continuous coverage that EXIST would provide 
gives relative immunity from the inevitable aliasing 
noise in scheduled pointed observations, and would allow 
the break frequencies to be measured with greater confidence. 

\subsection{Hard X-ray timing objectives for EXIST}
We next summarize what are perhaps the most compelling science 
goals for temporal surveys with EXIST. Each 
hard X-ray timing objective is described in 
only broad outline form; the details are being studied as part 
of the BHFP Mission Concept Study, and 
results will be reported elsewhere. These temporal surveys  
are listed in decreasing order of cosmic distance scale without 
implied corresponding order of interest or priority.

\subsubsection{\it GRBs at the limit}
This key objective of EXIST, which potentially has the 
best chance of locating and studying PopIII BHs giving rise 
to the very first gamma-ray bursts (GRBs) 
at z \about15, is of course not really in 
the category of a hard X-ray temporal survey of BHs (which may 
otherwise denote surveys or studies of 
persistent, or quasi-persistent, sources). The science 
goals for GRBs on EXIST are partly summarized by Grindlay et al 
(2003b). The GRB survey, and its neutron star cousin -- a survey 
for the most luminous superflares from SGRs in external galaxies 
(see below) -- have similar key attributes: 
both are possible since the EXIST survey 
and mission concept are designed to maximize sensitivity, FoV, and 
temporal coverage to enable detections of faint, rare events. The next 
objective is perhaps the highest priority example.

\subsubsection{\it Stellar disruption near supermassive BHs}
Stellar encounters with supermassive BHs (SMBHs), taken here 
to be with masses \ga10$^6$\Msun and so including even 
that in SgrA*, in the nuclei of 
galaxies are inevitable and indeed probably of fundamental 
importance to the buildup of SMBHs. A concise review of many 
of the developments in this rich subject is provided by 
Alexander (2003). For stars in the nuclei of galaxies with 
SMBHs with masses \la10$^8$\Msun, the tidal radius 
R$_t$ \about r(M/m)$^{1/3}$ 
at which stars of mass m and r are disrupted by the SMBH of 
mass M, lies outside the event horizon. Stellar debris from 
a star passing close enough to be disrupted on the first or 
subsequent periastron passages will be accreted by the SMBH 
and produce an energetic flare. Several such events were very likely 
detected in the ROSAT all sky survey as soft (kT \about0.1keV), 
luminous (\Lx \about10$^{42-44}$\lcgs) events as reviewed by 
Komossa (2002). Recent followup observations with HST 
(Gezari et al 2003) of particularly promising events 
in the galaxies RXJ1242.6-1119 and RXJ1624.9+7554, 
and with Chandra and XMM-Newton (Komossa et al 2004) 
of the RXJ1242.6 galaxy, have confirmed 
they are not in presently active Seyferts, thus further supporting 
the stellar disruption scenario. Finally, Wang and Merritt (2004) have 
incorporated the most recent M$_{BH}$ - $\sigma$ relation 
for BH masses from central velocity dispersions 
and calculated the rate of such events as 
\about10$^{-5}$ yr$^{-1}$ Mpc$^{-3}$ for non-dwarf galaxies (with 
the possibility of even higher rates in dwarf galaxies if they 
contain intermediate mass BHs), and Alexander and Hopman (2003) 
have shown that disruption events (as opposed to slow in-spiral 
events) are the dominant conributors to increasing the mass of 
the SMBH. 

Cannizzo, Lee and Goodman (1990) have modelled these stellar 
disruption events, finding they produce accretion luminosities 
at approximately Eddington values and with emission likely dominated by 
EUV and soft X-ray emission from the disk at effective temperatures 
\about10$^5$K. Their predicted accretion rate and thus 
luminosity remains roughly constant for \about1yr, followed by 
a powerlaw decay, L \about $\tau^{-1.2}$. Not calculated by Cannizzo et al 
(or anyone else, to our knowledge), but quite likely by analogy 
with other extreme transient accretion events (which these 
surely would be), is the possibility that in addition to the 
dominant soft thermal emission there is an underlying hard power 
law component. Stellar mass BHs in the Galaxy are usually first 
detected as ``soft'' X-ray transients but are  always 
accompanied by luminous hard power law emission (or Comptonized 
blackbody) which dominates the flux from the luminous thermal 
component at energies above \about10-20 keV (e.g., Tanaka 
and Lewin 1995). The luminosity in this hard component, broad 
band, is usually at least 10\% of the peak outburst luminosity. 
{\it Thus, if even just 1\% of the near-Eddington 
soft \Lx from tidal disruption of a star by a SMBH appears 
in such a hard tail, these  events will be 
readily detectable with EXIST.}

We estimate the detection rate as follows: using the Wang 
and Merritt (2004) disruption rate of 
\about10$^{-5}$ yr$^{-1}$ Mpc$^{-3}$ and a conservative (10$\sigma$) 
detection flux limit of \Fx(20-40 keV) \about 1 \X 10$^{-12}$ \fcgs, 
and ``typical'' (e.g. L$_*$) galaxy with central BH of mass 
10$^7$\Msun and space density \about0.02 Mpc$^{-3}$, and an 
assumed hard X-ray luminosity that is 
\Lx(\ga10keV) \about 0.01\Lx(soft), with 
\Lx(soft) \about L$_{edd}$ \about10$^{45}$ \lcgs, 
EXIST should detect these events out to \about100Mpc at a 
rate of \about30 yr$^{-1}$! An even more conservative assumption 
that the peak outburst luminosity, \Lx(soft), is instead only 
0.1L$_{Edd}$ brings this down to \about1 yr$^{-1}$. 

The recent Chandra and XMM observations of the RXJ1242.6 galaxy 
by Komossa et al (2004) show that the soft emission 
has declined even more rapidly, by a factor of \about200 over the 
intervening decade since the ROSAT observation. Most interesting 
of all is that the XMM spectrum shows the emission to be a 
power law (with photon index 2.5$\pm0.2$) detected out to 
at least 10 keV and with luminosity in the ROSAT band of 
\Lx(0.1-2.4keV) \about 5 \X 10$^{41}$ \lcgs. This provides at 
least indirect evidence that the disruption event itself does 
indeed include a hard component, as required if they are to be 
detectable with EXIST. In this case, also, it may be that the decay of 
the outburst is less than the observed (Komossa et al 2004) 
factor of \about200 since that is derived by comparing the thermal 
(outburst peak?) to non-thermal (decade later) components. If 
the non-thermal component were \about10\% of the thermal (instead 
of the 1\% value assumed above), then it would have decayed by 
a factor of \about20, or very close to the factor of 
\about14 = 9$^{1.2}$ predicted from the Cannizzo et al (1990) 
models for the power law decay over the 9 year interval 
between the ROSAT and XMM spectra.

Obviously many more such events must be observed. As noted 
in Komossa et al (2004), planned soft X-ray all sky monitor 
experiments (e.g. Maxi or Lobster) might do this but only for those 
galaxies with their central SMBH not obscured. EXIST would allow 
an unbiased survey of all such events, provided they do indeed 
produce a hard X-ray component. For encounters of sub-giants 
(rather than main sequence stars, as assumed above), the disruption 
is partial and the He WD core will eventually plunge into the 
SMBH after spiralling in. For these events, EXIST will have 
provided a trigger for LISA to then record the gravitational wave 
signature.

\subsubsection{{\it AGN and QSO variability and QPOs}}
We have already discussed the applicability of EXIST to 
the fundamental problem of measuring the variability spectra 
of AGN. EXIST would measure minimum variability timescales, 
which could be $\tau_{min}$ \about0.25d 
for emission from 3R$_{Scwarz}$ from a 
10$^7$\Msun SMBH. Given the corresponding flux limit (5$\sigma$) 
of 1mCrab = 1 \X 10$^{-11}$ \fcgs for a 0.25d or 4 orbit 
integration on any source, this would allow all AGN containing 
10$^7$\Msun SMBHs and accreting at 1\% of Eddington (thus 
with \Lx(\ga10 keV) \about10$^{43}$ \lcgs to be measured. These 
would be detected out to distances d \about100 Mpc. Assuming 
such Seyferts constitute about 1\% of the galaxy distribution, 
there should be some 80 Seyferts in this lowest mass (10$^7$\Msun) 
AGN sample for which power spectra could be measured. 
The numbers increase for more massive SMBHs if they are 
accreting at the same 1\% of Eddington rate, despite the 
fall off in their host galaxy numbers (with increased 
mass to accomodate the more massive SMBH) since the available 
detection volume increases faster.  QSOs with 
\Lx(\ga10 keV) \about10$^{45}$ \lcgs and SMBH masses 
\about10$^9$\Msun (again assuming 1\% Eddington) could be 
detected on their presumed minimum timescale 
$\tau_{min}$ \about25d  out to z \about0.8, 
or approximately where obscured AGN 
seem to peak in local number density. Thus a large sample of 
AGN and QSO light curves and power density spectra should 
be achieved from which to test models for QPOs and, more 
fundamentally, to constrain spin of the SMBHs if their mass 
is inferred independently (e.g. from reverberation mapping) 
by in effect measuring their innermost stable orbit radii as 
the radius appropriate to $\tau_{min}$.

\subsubsection{{\it Stellar mass BH variability and QPOs}}
Again, we have partly discussed the stellar mass BH case, 
where RXTE has done so much. The high background due to the 
large FoV of EXIST-RD makes it competitive only above \about20 keV 
and then only for bright (\ga100mCrab) sources. However there 
is physics to be learned by signficantly extending the energy 
range of variability and QPO studies. Determining $\tau_{min}$ 
over a significantly broader energy band can constrain lags 
and Comptonization models. Most unique, however, would be 
the first comprehensive survey of the evolution of QPOs 
with \mdot and spectra. That is, how do QPOs change their 
``spots'' when input parameters (e.g. \mdot) or output 
parameters (e.g. spectral shape and thus models) change? 
The migration of QPO peaks and power spectra slopes and 
breaks with input/output parameters will require the 
continuous coverage EXIST would provide; sporadic pointings 
will not track the full behaviour. As with AGN, this could 
lead to qualitatively new constraints on BH spin or, at the 
very least, the underlying physics of QPOs.

A key area for stellar BH timing is of course the detection and 
study of new (as well as recurring and known) transients. The 
imaging ASM nature of EXIST is required for discovery and 
locations of faint transients (e.g. \about10mCrab) in 
crowded fields (e.g. the central Bulge). The hard X-ray response 
is similarly critical for these heavily absorbed regions. The 
recent hard X-ray images of the galactic center region from INTEGRAL 
(e.g. Revnivtsev et al 2004) indicate a population 
of absorbed sources for which timing (and 
spectral) studies could elucidate their BH (vs. NS) nature. 

The imaging of EXIST has two further advantages for timing on 
relatively faint (for fast timing: \about 10-100mCrab) sources 
in that fluctuations (power spectra, etc.) can and would be 
measured simultaneously for background events as well as all 
other actual sources in the FoV. Coded aperture imaging does not 
scramble temporal signals or signatures, regardless of how 
many sources (and background) are in the FoV. 

\subsubsection{{\it Neutron stars and SGRs}}
Neutron star timing with EXIST will be optimal for pulsars 
(both accretion and rotation powered), with their hard spectra 
which extend out to \ga20 keV in both cases. The 
full sky each orbit imaging-ASM 
nature of EXIST, at high sensitivity per orbit, will facilitate 
the discovery of faint pulsar transients (e.g. Be-HMXB systems). 
As with galactic BHs, the study of QPOs from bright LMXBs 
containing NS primaries is competitive with the PCA or any 
planned 2-10 keV followon XRT mission by providing the 
diagnostic coverage above 20 keV. Indeed, one of the many 
attractive ``finder telescope'' features of EXIST (not just 
for timing) is that it could provide the hard X-ray coverage 
continuously and at no impact and thus allow the XRT to be a 
large area low mass/low resolution X-ray optic with sensitivity 
only below 7 keV (for Fe line coverage).

A unique advantage of EXIST for timing studies of NSs is to 
extend our study of soft gamma-ray repeaters (SGRs), or 
magnetars, to galaxies well beyond our Local Group. Whereas the 
normal SGR bursts from the 4 systems known (3 in the Galaxy, 
1 in the LMC) have \about1sec duration and \about30 keV exponential 
spectra with peak luminosities of typically \about10$^{41}$ \lcgs, 
two of the sources have shown ``superbursts'' or giant flares with 
shorter duration (\la0.1sec) initial spikes which reach peak 
luminosities \about10$^3$ times the normal bursts. Such superbursts, 
with \Lx(\ga10keV) \about10$^{44}$ \lcgs, could be seen with 
EXIST (for which the limiting flux is \about2 \X 10$^{-9}$ \fcgs 
for a 0.1sec detection)  out 
to the \about16Mpc or the distance of the Virgo cluster. 
Thus a sample of nearly 10$^3$ galaxies could be 
continuously sampled for super-burst SGRs. Given that the two 
known super-SGR bursts (in the Milky Way and LMC) 
have each occurred once in about 20y, 
a plausible detection rate of super-SGR events could be as 
high as \about50 yr$^{-1}$. This would allow the first survey for 
magnetars in external galaxies and enable tests for their formation 
sites (e.g. presumably in spiral disks), activity cycles and 
superburst production.

\section{Possible modifications to the Reference Design}
We mention a few of the ongoing studies for the BHFP that could 
further enhance EXIST as a hard X-ray timing mission, with even 
more capabilities appropriate to a next-generation X-ray timing 
mission. \\

\ni{\bf Energy range:} First, the energy band covered and in 
particular (for XRT objectives) the lower energy limit E$_{min}$  
is being studied. If E$_{min}$ could be reduced from 10 keV to 5 keV, 
the sensitivity to QPOs could be increased since signal count rates 
would double (for typical spectra approximated as power laws with 
photon index 2). However backgrounds would also increase, and by 
even larger factors if the FoV is maintained at the EXIST-RD value.
Thus  E$_{min}$ and FoV are correlated study options. 
Decreasing  E$_{min}$ also requires consideration of the detector 
and readout technologies since noise considerations as well as 
mass overburden make \about5keV thresholds difficult to achieve 
with the baseline CZT array technology. One option being considered 
is to have a portion of the CZT array apportioned to a lower energy 
E$_{min}$ detector system with total area perhaps 1/4 of the total 
available (to maintain comparable signal to noise with the primary 
CZT system).

The upper energy limit,  E$_{max}$, is also a primary design 
driver (which also affects  E$_{min}$ of course) and is being 
studied for optimum science and impact on mission cost and 
complexity. E$_{max}$ = 600 keV is the value for EXIST-RD, but 
values as low as 300 keV (cf. Table 1) are being studied. \\
 
\ni{\bf Field of view:} Decreasing the FoV would of course greatly 
reduce backgrounds and, for the same detector area, increase 
timing sensitivities. It impacts the primary survey objectives, and 
all sky monitor capability, by reducing exposure time per source 
correspondingly. The present large FoV is chosen to be that 
value where diffuse sky flux is comparable to (but larger than) 
detector backgrounds due to shield leakage and other processes. 
Other options are possible, such as smaller FoV at lower energies 
(e.g. \la30 keV) where the effects of bright point sources are 
largest. One option being studied is to reduce the FoV in the 
cross-scan direction (to maintain exposure time in the scan 
direction) at low energies with an interposed thin slat collimator. 
Other options are also being investigated. \\

\ni{\bf Pointing vs. scanning:} The EXIST-RD has a simple 
mission model: the telescope stack is zenith pointing and scans 
the full sky around the orbit. The pointing is very forgiving 
(\about0.5\deg) but the aspect requirements are tight (5\arcsec). 
The S/C control systems to do this can then also allow inertial 
pointing on a fixed target (or many targets, given the very 
large FoV). This would increase the exposure time per source or 
field that is achieved each orbit by a factor of (typically) 2, 
given the large FoV and thus exposure time per orbit in scanning 
mode. Nevertheless it is of interest for certain timing objectives 
and could easily be accomodated. \\

\ni{\bf Re-configurable telescope array:} The EXIST-RD has its 
three telescopes fixed, in the tower-stack 
configuration (Figs. 1 and 2) that allows 
each to view \about1/3 of the 180\deg~ fan beam unocculted by the 
others. Alternatively, if the three telescopes were mounted on 
the same plane (on a long bench; which would require a higher 
launcher shroud (Fig. 3) to accomodate and was therefore not chosen as 
the baseline), and they were each allowed to move in elevation 
(only), the three could be co-aligned to increase the available 
area by a factor of 3 for fixed pointing or slow scan pointing 
on a given source or group of sources. This would have the largest 
impact on timing objectives (particularly if combined with the 
smaller low energy FoV option) and allow (for the current design) 
\about4.5m$^2$ to be pointed on a given field. This option 
would entail additional mission complexity and (probably) mass, 
for the larger envelope. However tradeoffs can be considered 
as part of the current Study for the BHFP.  

\section{Conclusions}
EXIST in its current Reference Design configuration already 
incorporates several key attributes of an optimized hard X-ray 
timing mission. The X-ray Timing mission or observatory 
concepts discussed as the needed successor for RXTE are 
based in large part on the requirements to resolve individual 
pulse trains in QPOs from galactic BH and NS accretors. EXIST 
would not do this in its Reference Design but could, in principle, 
with some of the options being studied now by the EXIST team 
for the BHFP Study. However EXIST, in its primary survey mode 
configuration is already a superb ASM and would have sensitivity 
to conduct pathfinder science on the (much) longer timescales 
needed for variability studies of galactic BHs and AGN. Break frequencies 
in the power spectra of AGN, and QPOs in AGN, could be measured 
by EXIST over a range of interesting BH masses.

Of perhaps the most interest for novel timing studies of SMBHs 
is the suitability of EXIST to provide an unbiased (by low energy 
extinction) survey for stellar disruption near SMBHs that almost 
certainly lurk in bulges of all galaxies, without requiring 
these be known or detected as persistent sources by virtue of 
being AGN. Dormant AGN, or long since extinct AGN, are available 
for this study in sufficient numbers that surprisingly large 
rates of detection are predicted. More detailed study is needed 
to assess whether a disruption event will indeed produce 
an accompanying hard X-ray spectral component (as with galactic 
BH transients), but the recent XMM detection (Komossa et al 2004) 
of such a component in the faded ``afterglow'' of a 1992 likely 
stellar disruption discovered with ROSAT makes the hard X-ray 
component even more likely -- particularly if this XMM spectrum 
is later observed to decay with the $\sim\tau^{-1.1}$ possibly 
expected.


\begin{theacknowledgments}
I thank the EXIST Science Working Group for discussions and 
Roger Blandford and Martin Elvis for discussions of 
AGN variability and stellar disruption. I thank 
Bill Craig, Neil Gehrels, Fiona Harrison, Jaesub Hong, 
Ron Ticker and GSFC engineers for 
their work on developing the EXIST Reference Design. 
This work was partly supported by NASA grant NAG5 5-5396.

\end{theacknowledgments}



\section{References}
\begin{description}
\item Alexander, T. 2003, to appear in Coevolution of Black Holes 
and Galaxies, L.C. Ho, ed. (Pasadena: Carnegie Observatories)  
\item Alexander, T. and Hopman, C. 2003, ApJ, 590, L29
\item Gezari, S., Halpern, J., Komossa, S., and Leighly, K. 2003, 
\apj, 592, 42
\item Grindlay, J.E. et al 1995,  Proc. SPIE, 2518, pp. 202-210.
\item Grindlay, J.E. et al 2001, AIP Conf. Proc., 587, pp. 899-908
\item Grindlay, J.E. et al 2003a, Proc. SPIE, 4851, pp. 331-344
\item Grindlay, J.E. et al 2003b, AIP Conf. Proc., 662, pp. 477-480
\item Grimm, H.J., Gilfanov, M. \& Sunyaev, R. 2002, \aap, 391, 923 
\item Harmon, C.A. et al 2004, \apj, in press
\item Kaaret, P. 2004, these proceedings
\item Komossa, S. 2002, Rev. Mod. Astron., 15, 27
\item Komossa, S., Halpern, J., Schartel, N., Hasinger, G.,
  Santos-Lleo, M. and Predehl, P. 2004, \apj, 603, L17
\item Markowitz, A. et al 2003, \apj, 593, 96
\item McClinock, J. and Remillard, R. 2003, to appear 
in Compact Stellar X-ray Sources, eds. W.H.G. Lewin 
and M. van der Klis (astro-ph/0306213) 
\item Revnivtsev, M. et al 2004, \aap, in press (astro-ph//0402416)
\item Rodriquez, J. et al 2002, \aap, 386, 271
\item Shaw, S.E. et al 2004, \aap, submitted 
(astro-ph/0402587)
\item Tanaka, Y. and Lewin, W.H.G. 1995, in X-ray Binaries, 
W.H.G. Lewin, J. van Paradijs and E.P.J. van den Heuvel, eds. 
(Cambridge Press), pp. 126-174
\item van der Klis, M. 1998, in Many Faces of Neutron Stars, R. 
Buccheri, J. van Paradijs, and M. A. Alpar, eds. (Kluwer), p. 337
\item Voges, W. et al 1999, \aap, 349, 389
\item Wang, J. and Merritt, D. 2004, \apj, 600, 149

\end{description}

\end{document}